# Field reconstruction of holograms for interactive free space true three dimensional display

Guangjun Wang[a,*]

[a] BOWEI INTEGRATED CIRCUITS CO.,LTD., The 13th Research Institute of China Electronics Technology Group Corporation, Shijiazhuang, Hebei 050200, P.R. China

## Abstract

Holographic display can show pictures in the way that looks like the same with the real world, and is thought as the ultimate display technology. But due to the complexity of the traditional holographic technique, it can hardly be applied to commercial display systems. In this paper, the field reconstruction of holograms (or real-time reconstruction of holograms) method, which is more practical and better than traditional holographic technique in image quality, is first proposed for realizing the display of true 3D image at free space. By combining the advantages of geometrical optics and the wave optics, a true 3D display is successfully designed. And by designing an interactive image generation chip, the designed 3D display allows user to interact with the 3D image in the style that's natural to them.

## 1. Introduction

We often see three dimensional (3D) display technologies in science fiction

Email address: cetc2016@163.com

movies. This kind of magical displaying form always makes us admire. Nowadays, the 3D display technology attracts great academic and industrial attention due to its rapid development and applications in providing more realistic, natural, and extra depth images that are lacking in the traditional 2D display[1]. To achieve natural and comfortable 3D visual perception, a considerable amount of efforts have been dedicated, and some practical designs have been proposed. Among them the 3D display technology those based on two eyes parallax are the most mature ones[2]. But these kinds of 3D displays are not real 3D displays which are easily cause visual fatigue because of the accommodation–vergence conflict[3, 4] and they have some other disadvantages, i.e., the fixed range of the viewing distance, view position, and the fixed interpupillary distance value (usually 65mm). The parallax barrier technology seriously reduces the brightness of images, so it is replaced by lenticular sheet technology in most products. In the lenticular based autostereoscopic display, the role of the lenticular lens is to magnify and transfer the information of specific pixels to a designated position[5]. The depth-reversal is considered as an inherent disadvantage of the lenticular based or the parallax barrier based autostereoscopic displays. Integral imaging is a kind of true 3D technology, which is based on the principle of reversibility of light[6, 7]. But due to the poor resolution and complex structure, even it has been proposed by Lippmann for around a century, it still not been accepted by the market. Volumetric 3D display is another kind of true 3D displays, which also has some inherent defects[8]. The mechanical rotating components of it limits the image's spatial and brings in some potential danger.

What's more the resolution is reduced seriously at the central of the display.

It is difficult to construct a three-dimensional image in free space allowing view from all directions or position. From a mathematical point-of-view, the construction of 3D image in free space is a definite boundary solution problem of passive space. If the boundary condition (wavefront) is reconstructed the light field in the passive space will also be reconstructed. Then if we want to reconstruct a 3D image in free space, we have to obtain the control of boundary and control it at our will. In traditional hologram technology the recorded interferogram is used to set the "boundary condition" and thus we get a static 3D image[9]. If we want to display a dynamic 3D image the dynamic "boundary condition" is needed. In a typical computer-generated holograms system, spatial light modulator (SLM) usually used to generate the dynamic patterns [10]. But this kind of hologram display can't display large and clear 3D image due to the limitations of SLM[4]. A delicate design, proposed by Takaki, overcomes this defect to some extent[11]. In order to get a large computer generated hologram, a horizontal scanner is introduced to scan the elementary hologram generated by the anamorphic imaging system. The birth defect of computer-generated holograms is that it is difficult to provide a SLM with the pixel count high enough. A further difficulty is providing the data in real time. Calculating holograms usually require an immense calculation power.

The "boundary condition" (wavefront) reconstruction mechanism means we can't realize a full view 3D image because the excited boundary (screen or some other devices) always shades some view point. In fact, there is no need to achieve a full view 3D image display. Audiences usually sit or stand at one side of 3D display, so they don't care about the side of 3D images that can't be seen at their position. Based on these considerations, a new approach to display hologram was proposed by R. Häusslerur[12]. Instead of reconstructing the image that can be seen from a viewing region, the primary goal of him is to reconstruct the wavefront that can only be seen at a viewing window (VW). This holographic display omits unnecessary wavefront information and significantly reduces the requirements on the resolution of the spatial light modulator and the computation effort compared to conventional holographic displays.

Another thing need to be taken into consideration is how to interactive with the 3D image. This article describes a novel real 3D display method which is designed from the ground up. The novel design combines the advantages of geometrical optics and the wave optics. The main feature of this system is that it used the first proposed field reconstruction of holograms or real-time reconstructed hologram method to realize the display of 3D image. The novel system introduces the reference light and the virtual object light and let them impinge upon a screen, and then the interferogram is obtained on the screen. Just like the holographic technique, which can show depth information of scenery for that it has recorded the phase of light, in a field reconstructed holograms system, the depth information of scenery can also be provided by modulating the phase of light. So, the field reconstructed holograms can also be named as real-time phase modulating 3D image. And by designing an interactive image generation chip, the 3D display allows user to interact with it in the style that's natural to them. Although the devices used in this 3D display are all newly designed, similar system can be set up using commercial available components, which will make it easier to implement.

## 2. Design of the interactive true 3D display

There are few reports of true 3D displays with interactive ability. Thus, in order to realize a real 3D display with interactive ability, we have to design it from scratch. For a 3D display, an image generation component is necessary and if we can integral the interactive function into the component, a bi-functional system thus can be implemented.

**2.1 Design of interactive image generation chip**

Firstly, let's design a "T" type optical switch (TOS), which has three ports denoted as port A、B and C. As shown in Fig. 1a, the port A and port B are on a line and the port C located in between them and perpendicular to the line. The function of the TOS is that it can control the light incident from port A exit from either port B or port C. There are a lot of ways to realize such a TOS so there is no need to describe it in detail. And the following discussion will focus on the interactive free space real 3D display. Use several this kind of TOS, we can construct an optical switch chain (OSC): string several TOSs together and the let the port B of a TOS connecting to the port A of the TOS at its right side while the port A of the TOS connecting to the port B of the TOS at its left side. Then assemble a quarter wave plate followed by a polarization beam splitter at one end of the OSC and at the two unoccupied ports of the splitter connect a laser light source (make sure that the angle between the polarization direction of the light and the optical axis of the quarter wave plate is 45°) and a light

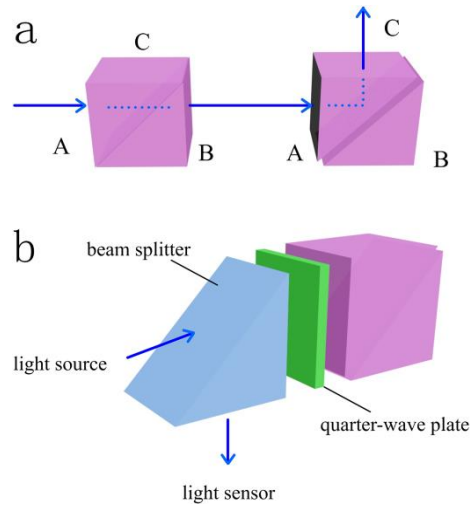

Fig. 1. Illustration of basic function of TOS. a) Two TOS in series b) a TOS connected with a quarter wave plate followed by a polarization beam splitter

sensor respectively, as shown in Fig. 1b. Then the light coupled into the OSC can exit through any port Cs of the TOSs in the OSC under control.

When the emitted light (from the OSC) hits on an object, some light will be reflected. The reflected light will go reversely along the original light path. The polarization direction of the reflected light will be rotated by ninety degrees when it transports through the quarter wave plate again. Under the action of polarization beam splitter, the reflected light finally hit the sensor, thus the information (such as finger touch action) brought back by the reflected light can be interpreted.

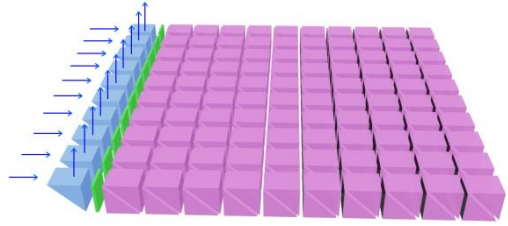

Fig. 2. Schematic of the OSA

Fig. 2. depicts a two dimensional optical switch array (OSA) which is made up of the above mentioned OSC. The function of the OSA is greatly extend when compare with the OSC. Since we can control the color, intensity and position of the output light of the OSA, it can be used to generate pictures. It can also be considered as a SLM (to generate image), but what different is that it can also be used as an image sensor and the two functions of the OSA can work simultaneously without interplay.

**2.2 Principle of field reconstructed holograms**

The configuration and principle of the proposed field reconstruction of hologram method are shown in Fig. 3 and Fig. 4 respectively. The system is analogous to a projection system. It contains an OSA, an imaging lens, a screen, a reference light source (denoted as R in the picture) and an eye tracking sensor. The OSA is used to generate the image and the imaging lens is used to project the image to free space and form a virtual object (the projected image). The screen is best to choose those made of transparent materials (such as a ground glass or holographic projection film). What different from a projection system is that the screen is no need to be placed at the image plain of the lens and the light source used here are coherent lights. Thus, rather than forming a bright spot on the screen, light from every point of the virtual object lit up a zone on the screen. Meanwhile, if the zone of the screen lit up by reference light overlap with the zone light by point light source on the virtual object, there will form an interferogram, which is also the hologram of the virtual object. The light emits from the virtual object can be written as:

$$O(x,y) = A_o(x,y)e^{i\varphi_o(x,y)} \tag{1}$$

And the reference light can be written as:

$$R(x,y) = A_r(x,y)e^{i\varphi_r(x,y)} \tag{2}$$

Then the amplitude and intensity of interferogram on the screen satisfies:

$$A(x,y) = O(x,y) + R(x,y) \tag{3}$$

$$I(x,y) = A(x,y) \cdot A^*(x,y) = A_o^2 + A_r^2 + 2A_rA_o\cos(\varphi_r - \varphi_o)$$

(4)

Where * denotes complex conjugate. For simplicity, the $\varphi_r(x,y)$ can be chosen as 0 and this is easy to implement under actual conditions. Then at the modulating action of $O(x,y)$, the transmitted part of reference light written

$$R_t = (T + \beta \cdot I(x,y)) \cdot R(x,y)$$

$$= (T + \beta \cdot (A_o^2 + A_r^2))R(x,y) + \beta \cdot A_r^2 \cdot O^*(x,y) + \beta \cdot A_r^2 \cdot O(x,y)$$

(5)

Where both T and β are constants. At proper condition T may negligibly. It is clear that the physical significance of these terms. The first term represents the attenuated reference light while the second term is the complex conjugate of $O(x,y)$, which means that it will generate a real image. And this real image can be seen by eyes at the conjugate position of the lens, as shown in Fig. 4. The third term has the same form of O(x, y), which will produce a virtual image. The difference between the real image and virtual image is that the former is a converged image while the latter is an emanative one. Thus the light from the real image converges to a certain position (the conjugate position of lens) and can be seen completely at this position. But the light from the virtual one diffuses to different directions so it can't be seen completely at any position. This phenomenon is depicted in Fig. 4 intuitively. Now you must have known the function of the eye tracking sensor. It is for the tracking of eyes position and helps to make sure that the eyes located at the watch window. From the above discussion we have understand the principle of field reconstruction of holograms method and know that by using this special method we can produce an image at free space. But because the light source is emitted from the plane surface of the OSA, the image generated on it is a two-dimensional graph. It is well known that lens can project a plain picture to another plain (image plain) and form a magnified image or a reduced image. Then even the field reconstruction of holograms method can translate an image to its conjugate image, it still can't convert a 2D picture into a 3D one. So we need to find other ways to obtain a 3D image. Since the field reconstruction of holograms method can display a real image at free space, in fact, it's quite easy to generate a 3D image with the help of this method. Put simply, the main function of the field reconstruction of holograms method is to convert a diffused real image (such as the image been casted to free space by a projection system, which can't be seen completely) to a visible converged image at its conjugate position (conjugate image plain). If the image plain of the projection system changed then the conjugate image plain will change, correspondingly.

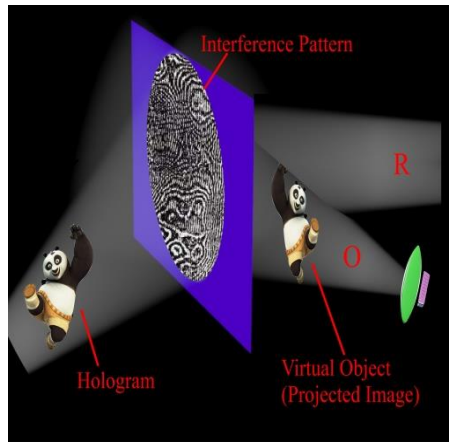

Fig. 3. The configuration of the proposed field reconstruction of holograms method. It contains an OSA (the pink component at right side), an imaging lens (the green component at right side), a screen (the blue square in the middle), a reference light source and an eye tracking component (not shown in the picture).

This way, the depth of field of field reconstructed holograms can be controlled by the projection system. As shown in Fig. 3, we can change the distance between the OSA and the lens to control the position of conjugate image plain (equals the scan of depth of field) to obtain a real 3D image at free space. A feasible method to realize the scan of depth of field is to make the OSA and/or the lens oscillate near a certain position.

After have solved the problem of displaying of 3D image, the next thing needs to be taken into consideration is how to interact with the obtained 3D image? The answer is: touch it with your fingers directly. When the 3D image is touched by finger, some of the light from the 3D image will be reflected by fingers. Due to the reversibility of the physical process, the reflected light will go through the screen, lens, OSA and finally received by light sensor. Then the light sensor can read the information (touch action) brought back by the light. Now the desired 3D image display and interactive functions are both realized in the designed system. It should be clearly noted that the designed 3D display system, if modified reasonably, can also be used to record high quality 3D image, but this is not the thesis of this paper so we don't discuss it in detail.

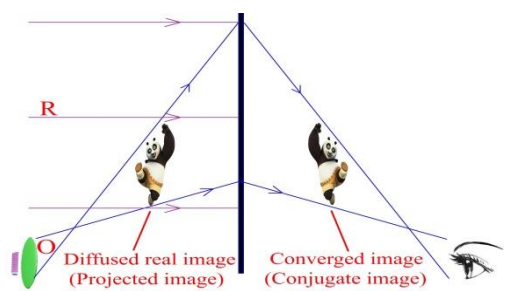

Fig. 4. Sketch that illustrates the principle of the field reconstructed holograms

The field reconstructed holograms can convert diffused real image (such as the image been casted to free space by a projection system, which can't be seen completely) to a visible converged image at its conjugate position (conjugate image plain). A 3D image thus can be displayed at free space when the image plain of the projection system changing.

## 3. Analyze and discussion

People, those don't familiar with optical physics, may doubt about the feasibility of the scan of depth of field scheme. Does it need large amplitude to acquire an acceptable display volume? Well, take mobile phones as an example. Even the moving distance of the charge coupled device (CCD) in a cellphone camera is much smaller than a centimeter (the thickness of a cell phone is only about 1cm), the camera can take photos clearly at a distance from one tenth meter to several meters. According to the reversibility of optical path, it means that using a similar system we can acquire a display volume of about several cubic meters even the oscillation amplitude of moving parts (image generation chip) is much less than a centimeter. What we need to do is to replace the CCD of a cellphone camera with an image generation chip, such as the above designed OSA, a digital mirror device (DMD) or some other things with the similar function. In order to realize a proper display space we need to configure the system reasonably. We can use the imaging properties of lens to analyze and design a proper system. The mean position (u) of image generation chip, the amplitude of oscillation of the image generation chip and the focal length (f) of the lens are all the important parameters need to be set delicately. Once these parameters are settled the properties of the system, the farthest display plain (denotes as v1) and the nearest display plain (denotes as v2) and the

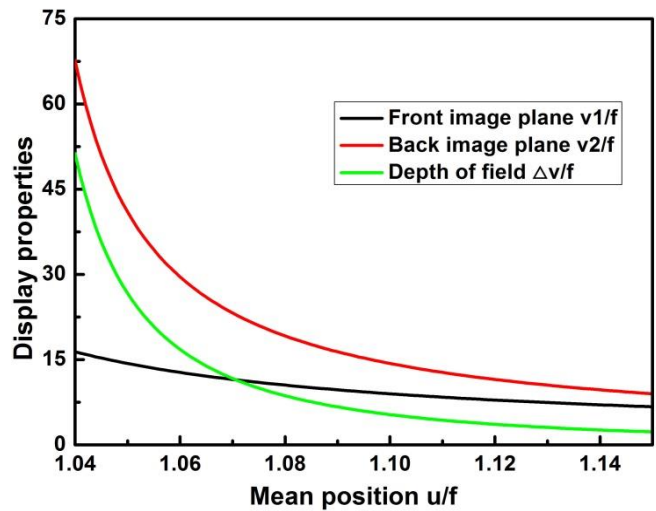

depth of field (v1-v2), are also been determined. As depicted in Fig. 5,

Fig. 5. Display properties of the system (the data is normalized with the normalized coefficient f)

even with the same oscillation amplitude (2.5 percent of the focal length of the lens), when the equilibrium point (mean position) of the image generation chip located at different distance from the lens, the effective display volume is varying. When the image generation chip moves from farthest position from the lens to the nearest place, the conjugate image plain will also moves from one plane to another plane. These two boundary planes can be called front image plane and back image plane and the zone between them is the effective display volume. As the mean point closer to the focal position of the lens both the front image plane and back image plane get further from the lens. Meanwhile the effective display volume, the zone between front image plane and back image plane, also gets bigger. The effect of amplitude on display volume is depicted in Fig. 6. It can be seen that enlarging the amplitude of the oscillation is another effective way to achieve a bigger display volume.

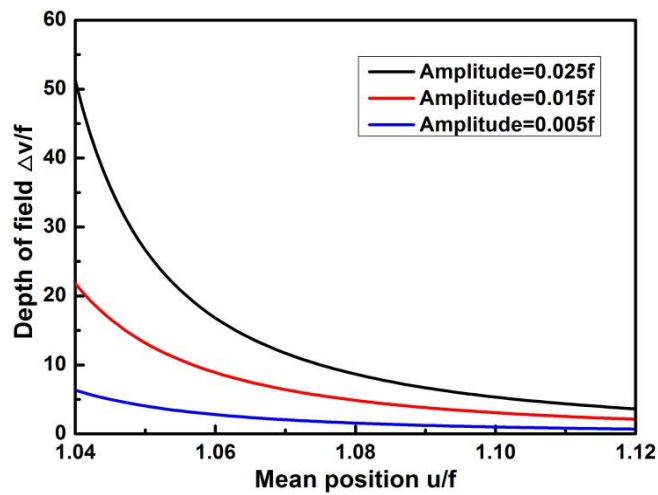

Fig. 6. The effect of amplitude on depth of field

Based on these principles, a proper image's spatial can be acquired by tunneling these parameters. An important information is that if a proper mean position has been chosen, a large display volume can be obtain even at the condition that the amplitude of the moving component in the system is negligible. Rather than using a moving chip, use a tunable lens can also realize the function of scanning of field depth. There are some other alternative ways, for example inserting a transparent media with controllable refractive index between OSA and lens. By controlling the refractive index of the transparent media we can also control the effective distance between the OSA and lens, and thus we can control the position of conjugate image plain. In this way we can obtain an all solid system without moving component.

It is different from a 2D displayer (i.e. project system) that the view point of this kind of 3D display is limited. Usually, difference means some advantages in some aspect and/or some disadvantages in other aspect. One of the birth defects of conventional projection system is that the brightness is very low, which makes it almost impossible to work outside. This is mainly because the light scattered by the scree attenuates as the inverse square of the distance from the screen. But this will not happen in the novel 3D display. The field reconstruction of hologram process makes sure that the diffractive light will converge at the watch window (also the conjugate position of project lens). Then there is no need to worry about the brightness of this novel 3D display system. Another benefit of the field reconstructions of hologram is that the generated holographic interferogram gets rid of the constraints of the resolution of photographic film. So it can obtain the clearest hologram. This in turn helps a lot in improving the quality of the image. What's more the image quality of traditional hologram is easily affect by speckles that generated by the random interference among object points[4]. And the recording condition of a conventional hologram is really strict, even the vibration caused by talking will cause the failure of the recording process. All of these inherent disadvantages of tradition holographic technique will not happen in the field reconstructed holograms technology. Meanwhile，the 3D image generation process of this special 3D display just like extruding a 2D pictures into a 3D ones. In a computer generated hologram system, the interferogram is calculated by computer, thus it would require an immense calculation power. And the more complex the displayed object the longer it takes. But in our system, the interferogram is generated by the direct interference of the reference light and the object light, and this physical process takes almost no time (no matter how complex the displayed object is). So, there is no need to generate the data for the interferogram. Thus, the amount of data needs to be processing almost the same with that in a traditional 2D display system. Hence, a commonly used mobile phone processor is quite qualified for the associate data processing and this in turn ensures the affordability of the system to a large extent. Another benefit of the system is that the visible solid angle can be very large. The visible solid angle is the solid angle that determined by the boundary of the screen and the watch widow ($\sim \frac{\text{the area of the screen}}{\text{distance from the screen to watch window}}$). Any image point inside the solid angle, no matter it is in front of the screen or behind the screen, is visible. The feeling of using such a 3D display is similar to that watching the real scenery through a window. So, the larger the screen the larger the visible solid angle can be obtained. This feature makes sure that enlarging the image's spatial will not increase the cost of the system obviously. For example，an screen about one square meter with a watch distance of one meter can be easily achieved, whose solid angle is around 1. In addition, even the viewpoint is limited in the above designed system, there are a lot of ways to extend it to a multi-view one.

## 4 Conclusion

In conclusion, it is first proposed that the field reconstruction of holograms method (or real-time reconstruction of holograms), which is practical and shares the same advantage of holographic technology. By combining the advantages of geometrical optics and the wave optics, a true 3D display is successfully designed. And by designing an interactive image generation chip, the 3D display allows user to interact with it in a direct way. The feasibility and properties' of the designed system were analyzed in detail. Even this article is focus on the 3D display, the proposed principle and methods have instructional significance for many other fields, such as taking 3D photos, augmented/virtual reality device and so on.

## Acknowledgement

This work was not supported by any fund.